\documentclass[conference]{IEEEtran}
\IEEEoverridecommandlockouts
\usepackage{import}
\usepackage{cite}
\usepackage{amsmath,amssymb,amsfonts}
\usepackage{algorithmic}
\usepackage{graphicx}
\usepackage{textcomp}
\usepackage{xcolor}
\usepackage{tabularx}
\usepackage{booktabs}
\usepackage{pifont}   
\usepackage{booktabs}
\usepackage{multirow}
\usepackage{gensymb}
\usepackage[most]{tcolorbox}
\usepackage{xcolor}
\usepackage{enumitem}
\usepackage{pbalance}
\usepackage{stfloats}

\renewcommand{\figurename}{Figure}
\newcommand{\cmark}{\ding{51}} 
\newcommand{\xmark}{\ding{55}} 
\newcommand{\pmark}{\raisebox{0.1ex}{\scalebox{0.7}{$\triangle$}}}   

\def\BibTeX{{\rm B\kern-.05em{\sc i\kern-.025em b}\kern-.08em
    T\kern-.1667em\lower.7ex\hbox{E}\kern-.125emX}}

\newcommand{\projectname}{\textbf{\texttt{INSIGHT-HPC}}}

\begin{document}

\title{INSIGHT-HPC: An Interpretable System Intelligence for Global HPC Telemetry%
\thanks{© 2026 IEEE.  Personal use of this material is permitted.  Permission from IEEE must be obtained for all other uses, in any current or future media, including reprinting/republishing this material for advertising or promotional purposes, creating new collective works, for resale or redistribution to servers or lists, or reuse of any copyrighted component of this work in other works.
}}

\author{\IEEEauthorblockN{Rupak Kadel}
\IEEEauthorblockA{\textit{Department of Computer Science} \\
\textit{Texas Tech University}\\
Lubbock, Texas \\
rupak.kadel@ttu.edu}
\and
\IEEEauthorblockN{Jie Li}
\IEEEauthorblockA{\textit{Department of Computer Science} \\
\textit{Texas Tech University}\\
Lubbock, Texas \\
jie.li@ttu.edu}
\and
\IEEEauthorblockN{Yong Chen}
\IEEEauthorblockA{\textit{Department of Computer Science} \\
\textit{Texas Tech University}\\
Lubbock, Texas \\
yong.chen@ttu.edu}
}

\maketitle

\begin{abstract}

Modern high-performance computing (HPC) clusters integrate multi-generation CPUs and GPUs, which generate rich, high-dimensional telemetry streams from out-of-band baseboard management controllers (BMC), job schedulers, and facility power distribution units. Detecting anomalies within this telemetry is critical to prevent energy waste and costly node failures. Existing approaches often rely primarily on statistical deviations and require stronger labels than actual production environments can provide. In addition, these detectors have a limited operational context for interpreting alerts. We present ~\projectname, a multi-plane anomaly-detection architecture that combines statistical, workload-aware, and physics-guided evidence into operator-interpretable episodes. \projectname~integrates heterogeneous, large-scale telemetry streams and transforms noisy minute-resolution observations into de-duplicated fault episodes with operator-facing explanations. This design allows the system to distinguish workload-explained behavior from genuine infrastructure anomalies while producing interpretable alert summaries for operators. On production HPC telemetry with weak operational labels and controlled HPC-relevant fault injections, \projectname~achieves a stronger precision–recall balance than all evaluated baselines. The system achieves a precision of 0.913, a recall of 0.817, an F1 score of 0.862, and an AUROC of 0.896. These results demonstrate the value of combining workload context, physical consistency, and explainable episode-level fusion.
\end{abstract}


\section{Introduction}
High-performance computing (HPC) systems form a foundational part of scientific infrastructure. These systems are critical for computationally intensive workloads such as engineering design, nuclear physics,  materials discovery, computational biology, climate modeling, and large-scale AI workloads~\cite{aksar2023prodigy,xia2025effective}. HPC sites face the challenge of maintaining high utilization from thousands of components while also managing dense telemetry streams, aggressive power envelopes, and cooling constraints that directly affect availability, efficiency, and performance~\cite{netti2021conceptual, bautista2019collecting}. At this scale, the cost of an undetected anomaly is not just a failed job. Instead, correlated hardware degradation and environmental stress can cascade across entire racks and can even shorten component lifetimes~\cite{sankar2011impact, han2021depth}. The SC'21~\cite{shin2021revealing} characterization of Summit's 200 PF pre-exascale facility across 4,626 nodes and 840,000 jobs demonstrates that power-thermal dynamics at this scale exhibit rich non-stationary structure that cannot be fully anticipated at design time.

Contemporary HPC nodes expose telemetry through a layered stack of monitoring interfaces. Out-of-band baseboard management controllers (BMCs) with Redfish streaming provide per-second readings of CPU temperatures, fan RPM, inlet and return air temperatures, and system-level power draw. Workload schedulers, such as Slurm, expose per-job CPU utilization, memory usage, and job state via Slurm REST API. Facility-side smart power distribution units (PDUs) contribute per-outlet power measurements, enabling cross-plane verification between node-reported and facility-measured power. Integrating these continuously generated sources presents a Big Data challenge because they differ in schema, arrival delay and sampling semantics. CPU and GPU nodes also expose different sensor sets. Modern HPC systems often use a scalable monitoring tool to aggregate these heterogeneous sources into minute-resolution multivariate time series (MTS) comprising multiple concurrent sensor metrics per node~\cite{li2020monster}. This rich set of metrics can reveal subtle anomalies, but distinguishing genuine faults from legitimate workload dynamics remains challenging.

Many HPC anomaly detectors are siloed within a single telemetry plane. This narrow focus often misses critical signals that only surface through cross-plane comparison or when thermal behavior deviates from its expected relationship with workload and power~\cite{chen2018detecting}. Lacking this physical grounding, these models can confuse standard workload transitions with genuine hardware faults. Residual-based physical modeling has long been a standard answer to this problem in fault detection and isolation~\cite{isermann2005fault,hwang2009survey}, but it has seen only limited use in HPC anomaly detection. Explainability is also still limited. Existing systems can often point to features that contributed to a score within a single model~\cite{aksar2023prodigy}. However, they usually cannot explain which kinds of evidence drove a final alert, whether that evidence is more consistent with a workload effect or a hardware problem, or what an operator should do next.

To address these gaps, we present \textbf{\projectname} (An \textbf{IN}terpretable \textbf{S}ystem \textbf{I}ntelligence for \textbf{G}lobal \textbf{H}PC \textbf{T}elemetry), a novel multi-plane anomaly-detection architecture that progressively refines statistical deviations using workload context and physical consistency before episode-level fusion. Our contributions are as follows:
\begin{enumerate}
\item We design workload-conditioned anomaly reasoning using Slurm job boundaries, workload profiles, and peer-node behavior.
\item We introduce physics-guided validation using power, thermal, utilization, and cross-plane consistency.
\item We evaluate \projectname~on 79 days of NSF-funded REPACSS (\textbf{RE}motely-managed \textbf{P}ower \textbf{A}ware \textbf{C}omputer \textbf{S}ystems and \textbf{S}ervices) production telemetry with 29 operator-recorded failures and 140 controlled fault injections, achieving an F1-score of 0.862.
\item We generate tiered, auditable alerts with job context, violated constraints, dominant signals, and model attribution.
\end{enumerate}

\section{Related Work}
\begin{table*}[!t]
\centering
\caption{Capability comparison of \textsc{\projectname~} with prior HPC anomaly detection systems.}
\label{tab:taxonomy_comparison}
\small
\renewcommand{\arraystretch}{1.15}

\resizebox{\textwidth}{!}{%
\begin{tabular}{lcccccccc}
\toprule
\textbf{Capability} 
& \textbf{RUAD} 
& \textbf{Prodigy} 
& \textbf{Refine} 
& \textbf{NodeSentry} 
& \textbf{GRAAFE} 
& \textbf{ThermADNet} 
& \textbf{DeepHYDRA} 
& \textbf{\textsc{\projectname~}} \\
\midrule

Job-aware modeling 
& \xmark & \pmark & \xmark & \cmark & \xmark & \xmark & \xmark & \textbf{\cmark} \\

Peer-node awareness 
& \xmark & \xmark & \xmark & \pmark & \cmark & \pmark & \cmark & \textbf{\cmark} \\

Cross-plane telemetry reasoning 
& \xmark & \xmark & \xmark & \xmark & \xmark & \cmark & \xmark & \textbf{\cmark} \\

Physics-guided reasoning 
& \xmark & \xmark & \xmark & \xmark & \xmark & \pmark & \xmark & \textbf{\cmark} \\

Explainability 
& \xmark & \cmark & \xmark & \pmark & \xmark & \cmark & \pmark & \textbf{\cmark} \\

Operator-facing alert tiers 
& \xmark & \xmark & \xmark & \pmark & \xmark & \xmark & \xmark & \textbf{\cmark} \\

\bottomrule
\end{tabular}%
}

\vspace{2pt}
\footnotesize{
\textbf{Legend:} \cmark~Supported \quad
\xmark~Not supported \quad
\pmark~Partial support (limited scope or integration)
}
\end{table*}

\subsection{Supervised Anomaly Detection}
Early ML-based HPC anomaly detectors were mostly supervised. Tuncer et al.~\cite{tuncer2017diagnosing} used statistical features from hardware counters and resource-usage time series with classifiers such as kNN, decision trees, AdaBoost, and random forests. Netti et al.~\cite{netti2020machine} followed a similar supervised direction by combining FINJ-based fault injection with classifiers such as SVM and random forests on the Antarex cluster. GRAAFE~\cite{molan2024graafe} uses graph neural networks to predict transitions from normal operation to anomalous states. However, it is difficult to maintain the labeled samples in production HPC clusters, where new hardware or workload behavior can introduce new anomaly patterns.

\subsection{Semi-supervised Anomaly Detection}
Semi-supervised methods reduce the labeling burden by assuming that some of the training data can be labeled. Borghesi et al.~\cite{borghesi2019anomaly} proposed a semi-supervised autoencoder approach in which the model learns normal node behavior and detects deviations via reconstruction error. Proctor~\cite{aksar2021proctor} uses autoencoder-based representation learning and then applies a supervised classifier for anomaly diagnosis. Aksar et al.~\cite{aksar2022albadross} later introduced ALBADross, an active-learning-based anomaly diagnosis framework that reduces the number of labeled samples required to diagnose known performance anomalies. ThermADNet~\cite{ardebili2026elevating} explicitly uses a semi-supervised learning approach for datacenter thermal anomaly detection. DeepHYDRA~\cite{stehle2024deephydra} is a hybrid approach that combines DBSCAN with learning-based anomaly detection and is often described as semi-supervised because its training includes a supervised loss component. These systems are more realistic than fully supervised methods, while still depending upon some form of clean normal data or labeling.

\subsection{Unsupervised Anomaly Detection}
Unsupervised methods are attractive in production HPC because they avoid requiring dense anomaly labels. RUAD~\cite{molan2023ruad} remains mostly node-centric and does not use Slurm job context, so it cannot directly separate workload-driven changes from infrastructure-side faults. Prodigy~\cite{aksar2023prodigy} provides job- and node-level analysis for operators, which makes it one of the strongest prior HPC baselines for explainable anomaly detection. Refine~\cite{sencan2025refine} is a robust unsupervised HPC anomaly detection framework designed to handle contaminated training datasets, which directly addresses a weakness of many reconstruction-based systems. USAD~\cite{audibert2020usad} is an explicitly unsupervised adversarial autoencoder for multivariate time-series anomaly detection. NodeSentry~\cite{xia2025effective} treats job-boundary segmentation as part of its anomaly-detection pipeline. However, it remains mainly workload-aware and does not explicitly validate anomalies using physical consistency checks. 

\textbf{Positioning of \projectname:} Despite this progress, several gaps remain. Most prior systems operate within a single telemetry plane. Job context and physical consistency checks also remain underused in HPC anomaly detection, while explainability is often limited. \projectname~addresses these gaps through a hybrid design. Table~\ref{tab:taxonomy_comparison} summarizes its positioning relative to prior work.

\section{Methodology}
\begin{figure*}[!t]
\centering
\includegraphics[width=\textwidth]{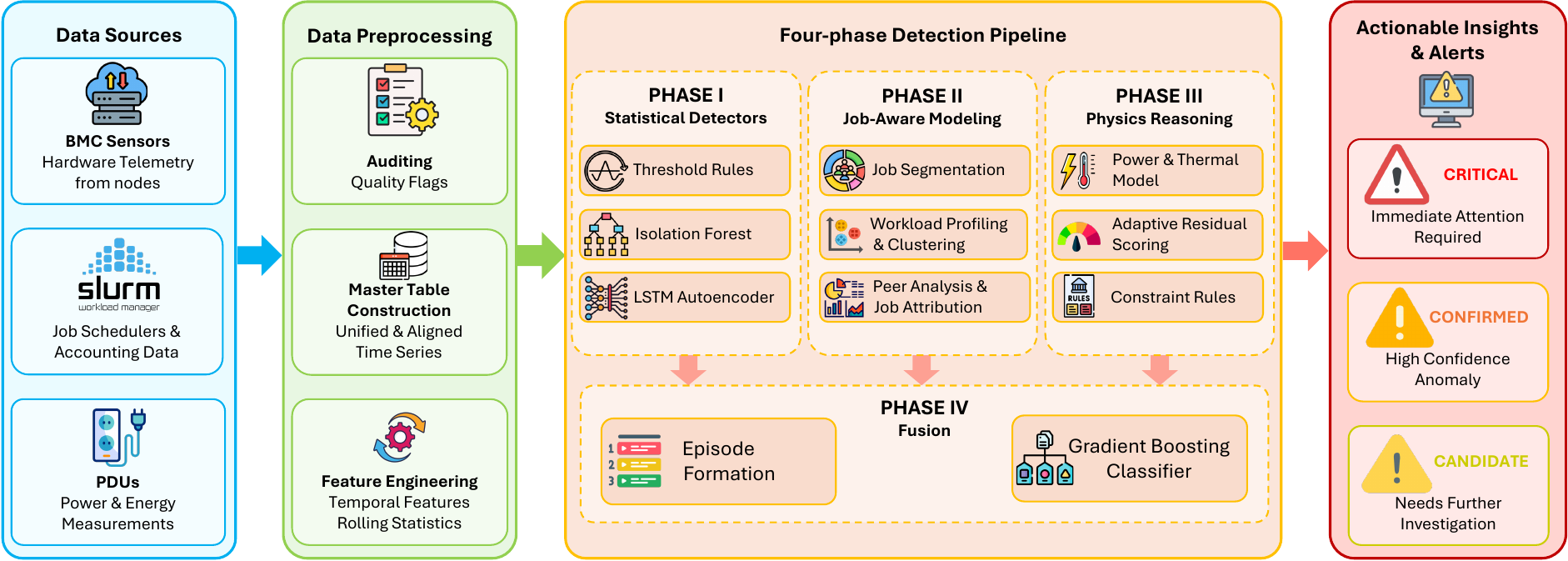}
\caption{The high-level system architecture of \projectname. The system ingests telemetry from multiple sources, aligns them into a per-host table, and derives features from them. These features are fed into Phases I--III, whose outputs are fused in Phase IV into candidate episodes and emitted as operator-facing alerts.}
\label{fig:architecture}
\end{figure*}

The overall system architecture of \projectname~is shown in Figure~\ref{fig:architecture}. The system ingests a rich set of metrics from multiple sources, such as BMC sensors, Slurm, and PDUs, which are then preprocessed and feature-engineered. The detection pipeline comprises \textit{four} phases. Phase I flags statistical outliers relative to each node's history. Phase II runs alongside Phase I, detecting behavior unusual for the current workload. Phase III then interprets the resulting anomalies through physics models and constraints. Finally, Phase IV fuses the statistical, workload-conditioned, and physics-based evidence into tiered, attributable alerts that operators can act on.

\subsection{Data Preprocessing}
Raw HPC telemetry is not directly usable for anomaly detection because sources report at different rates and timestamps may not align. As a result, data preprocessing is more than a cleaning step; it is the stage of the pipeline that builds a consistent, usable view of system behavior.

\subsubsection{Auditing}
The auditing stage marks suspicious raw sensor behavior before the anomaly detectors are trained. Rather than deleting rows immediately, we attach a per-sensor quality mask indicating whether the reading is a spike, flatline, or bad-range value. For spike auditing, we compute a rolling audit z-score (i.e., standard score) over a 120-sample window:
\begin{equation}
z_{audit} = \frac{v(t) - \mu_{roll}}{\sigma_{roll}}
\end{equation}
where $v(t)$ is the current reading, $\mu_{roll}$ and $\sigma_{roll}$ are the rolling mean and standard deviation, respectively. A spike flag is raised when \(z_{\mathrm{audit}} > 5.0\) and the absolute change ($\delta_{abs}$) exceeds a metric-specific floor. The absolute floor prevents nearly flat signals with very small variance from producing artificially large z-scores for harmless changes. Flatlines are detected by measuring the length of consecutive repeated readings. The default threshold is 480 minutes, though it is tightened for critical sensors such as temperature. For a bad-range detection, we use percentile-based expected bounds:
\begin{equation}
v(t) \notin [p_{0.001} * \beta_{lo},  p_{0.999} * \beta_{hi}]
\end{equation}
where $p_{0.001}$ and $p_{0.999}$ are the empirical $0.1^{th}$ and $99.9^{th}$ percentiles, $\beta_{lo} = 0.95$, and $\beta_{hi} = 1.05$. The 5\% buffer makes the range slightly more tolerant to normal drift, while still excluding clearly questionable sensor values.

\subsubsection{Master Table Construction}
Once the raw streams are audited, we build the master table, a per-node, per-minute table that combines BMC telemetry, Slurm context, and PDU signals on a common timestamp. Every job’s interval [start\_time, end\_time] is expanded to one row per minute so it can be joined directly to telemetry timestamps. If no job is active for a minute, the table records active\_job\_count = 0, indicating the node is idle. This is needed because raw telemetry alone does not indicate whether a power jump is suspicious or whether a large job has started. Then, we perform selective masking. For instance, instead of blindly removing all flagged data, we preserve spikes during active jobs as they may be legitimate workload effects. This is important because otherwise the training data would throw away exactly the workload transitions that later phases need to understand.

\subsubsection{Feature Engineering}
Each aligned node-minute is converted into features capturing \textit{short-term change}, \textit{longer-term drift}, \textit{missing-sensor behavior}, \textit{workload context}, and \textit{periodic context}, as summarized in Table~\ref{tab:features}.

\textit{Short-term change} features capture how each telemetry signal changes over a short period of time. For each sensor metric, we compute the rate of change, rolling means, standard deviations, and quantiles over trailing windows of 5, 15, and 60 minutes. These features allow the detector to distinguish stable behavior from sudden jumps. The \textit{longer-term drift} features calculate the longer-horizon linear drift slopes over day-scale windows for power and thermal metrics. These slopes are useful for detecting slow degradation, such as when a sensor may not appear abnormal at any given moment but may gradually drift away from its usual baseline. \textit{Missing sensor behavior} features handle missingness and sensor silence. For safety-critical signals such as fan speed, system power, CPU power, and temperature, we maintain counters that record the duration a value has been missing. This makes prolonged sensor silence visible to the detector. The next group of features adds \textit{workload context} from Slurm. For each node and minute, we encode whether the node is idle or allocated, whether multiple jobs share the node, and how recently the active job set changed. Finally, we add simple time-based \textit{periodic features} to help the models learn recurring daily and weekly patterns, such as typical work schedules or periodic cooling cycles. 

\begin{table}[t]
\centering
\caption{Example features computed in Feature Engineering.}
\label{tab:features}
\footnotesize

\resizebox{\columnwidth}{!}{%
\begin{tabular}{lccccc}
\toprule
\textbf{Feature Group} & \textbf{Example Features} \\
\midrule
\textbf{\texttt{Short-term change}}       & cpu\_power\_roc5, fan\_speed\_rmean15, etc. \\
\textbf{\texttt{Longer-term drift}}       & inlet\_temp\_slope12h, sys\_power\_slope24h, etc. \\
\textbf{\texttt{Missing-sensor behavior}} & fan\_missing\_streak, cpu\_power\_missing\_streak, etc. \\
\textbf{\texttt{Workload context}}        & active\_job\_count, is\_running\_job, etc. \\
\textbf{\texttt{Periodic context}}        & hour\_sin\_enc, dow\_cos\_enc, etc. \\
\bottomrule
\end{tabular}%
}
\end{table}

\begin{figure}[t]
\centering
\includegraphics[width=\columnwidth]{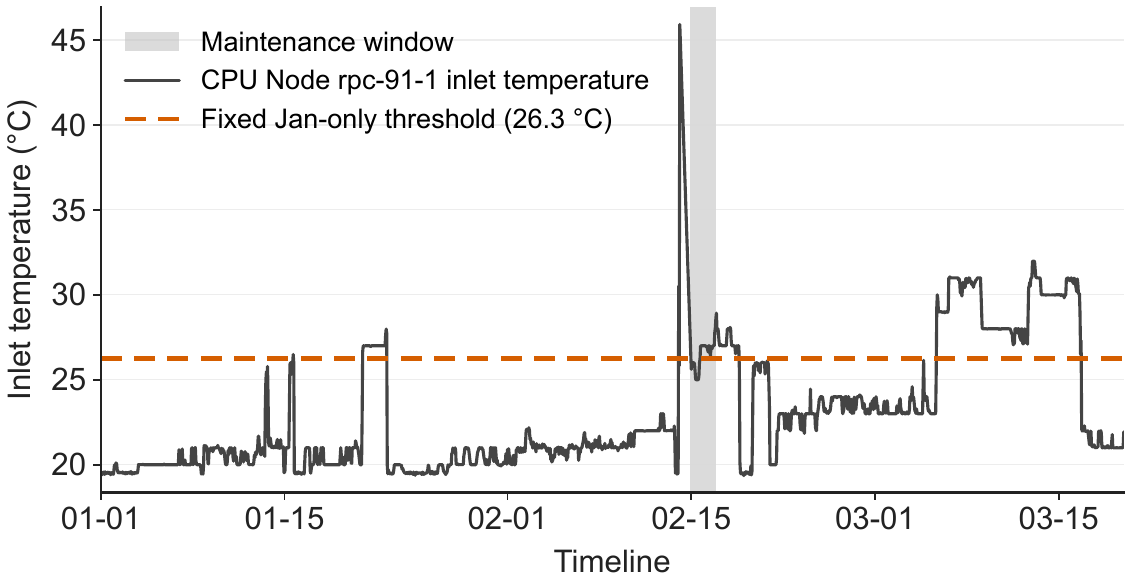}
\caption{Regime shift on a CPU node rpc-91-1 after maintenance window}
\label{fig:regime_shift}
\end{figure}

HPC telemetry also exhibits substantial regime-dependent variation over time. Maintenance operations, firmware updates, cooling changes, and evolving workload mixes can shift the normal operating baseline of a cluster even when no fault is present~\cite{goiri2015coolair,naug2024sustaindc}. As shown in Figure~\ref{fig:regime_shift}, the inlet temperature baseline of a CPU node rpc-91-1 changed noticeably after a maintenance window. Without accounting for such regime shifts, rolling statistics computed over long histories may incorrectly treat normal baseline transitions as anomalies.

To reduce this effect, we annotate each $(\text{node}, \text{timestamp})$ row with a regime identifier derived from available maintenance and operational metadata. All rolling-window and drift-based features described above are computed independently within each regime, preventing statistics from crossing regime boundaries. When regime metadata is unavailable, the identifier remains constant, and the pipeline naturally falls back to standard past-only windowing. This design allows the feature-engineering stage to incorporate operational context when available while remaining compatible with fully unsupervised deployments.

After feature construction, continuous features are normalized separately for each host using training-period statistics. We apply a minimum-variance floor and cap extreme normalized values to prevent unstable sensors from dominating later models. The final output is a feature-aligned table that keeps the physical meaning of the original telemetry while also making it easier for the statistical, job-aware, and physics-guided phases to reason over node behavior.

\subsection{Phase I: Statistical Detectors}
Phase I scans the feature-aligned master table for behavior that appears statistically unusual at the node and minute levels. It prioritizes coverage using several complementary detectors.

We use \textit{three} detector families because no single method is reliable across all the failure modes observed in HPC telemetry. \textbf{Threshold rules} provide the simplest baseline. They target obvious conditions like high system power, high CPU temperature, high inlet temperature, fan failure patterns, idle power surges, flatlines, and sensor dropouts. \textbf{Isolation Forest} identifies outliers by isolating observations, rather than profiling normal data points~\cite{liu2008isolation}. Its score is based on how quickly a point gets isolated by random tree partitions:
\begin{equation}
    s(x) = 2^{-\frac{E(h(x))}{c(n)}}
\end{equation}
with
\begin{equation}
    c(n) = 2H(n-1)-\frac{2(n-1)}{n}
\end{equation}
where $h(x)$ is the path length for point $x$, $E(h(x))$ is the expected path length across trees, $H(i)$ is the harmonic number, and $c(n)$ is the normalization constant representing the average path length of an unsuccessful search in a Binary Search Tree. We then employ the \textbf{LSTM autoencoder}, which is a sequence-to-sequence reconstruction model over 30-minute sliding windows~\cite{malhotra2016lstm}. It adds a temporal view by scoring whether a short telemetry window can be reconstructed as normal behavior. For each minute, we compute the mean absolute reconstruction error across all features:
\begin{equation}
    e_{recon}(t) = \frac{1}{F}\sum_{f=1}^F|x_f(t)-\hat{x}_f(t)|
\end{equation}
where $x_f(t)$ and  $\hat{x}_f(t)$ are the original and reconstructed values of feature $f$ at time $t$. Because windows overlap, we average the errors for each minute and normalize the result per node. To prevent learning anomalies as normal behavior, we clean the training set in two passes: first by removing windows flagged by Isolation Forest, and second by removing the top 1\% of highest-error windows.

The phase score is the mean of these three normalized detector outputs, and a node-minute is forwarded as candidate evidence only when this score remains above a fixed threshold. Because detector scores also shift with operating conditions, we convert each output to a per-host trailing rank against its recent history. This keeps firing rates more stable across regimes and emphasizes fresh deviations over persistent plateaus. Phase I therefore treats detector outputs as screening evidence, not final fault decisions, forwarding unusual node-minutes for later workload and physical validation.

\subsection{Phase II: Job-Aware Modeling}
Phase II adds workload context and cross-node comparisons to distinguish genuine faults from workload-driven changes in power, temperature, and utilization. Running in parallel with Phase I, it reconstructs the job context for every minute and compares nodes sharing a workload. The detailed description of each stage in this phase is as follows:

\subsubsection{Job Segmentation}
We begin by segmenting each node’s timeline using Slurm job boundaries. The per-minute feature-aligned master table produced during preprocessing is the right starting point for statistical detection, but workload-aware reasoning needs a larger unit than an individual timestamp. For that reason, we reduce the stream into contiguous segments during which the active job set on a node remains unchanged. A new segment begins when the primary job changes, when the set of active jobs changes, or when a sufficiently large telemetry gap appears. For each host $h$, a boundary is declared at minute $t$ when:
\begin{equation}
\begin{split}
    b_h(t) \;=\; & \mathbf{1}\!\big[\, j_h(t) \neq j_h(t{-}1) \,\big] \\
    &\vee\; \mathbf{1}\!\big[\, S_h(t) \neq S_h(t{-}1) \,\big] \\
    &\vee\; \mathbf{1}\!\big[\, t - t_{prev} > 120\,s\,\big]
\end{split}
\end{equation}
where $j_h(t)$ is the primary job ID, $S_h(t)$ is the sorted set of active job IDs, and the third clause closes a segment across telemetry gaps exceeding two minutes. Each segment is annotated with basic operational properties such as $is\_idle$, $is\_multi\_job$, $duration\_min$, and the total requested CPU count for the respective job IDs. Very short segments are excluded because they do not provide enough samples to build a stable workload profile.

\subsubsection{Workload Profiling and Clustering}
Each non-idle segment is then reduced to a compact behavioral profile. Instead of keeping a full minute-by-minute trace, we summarize the segment into roughly 30 engineered features that capture the workload’s physical signature. These include power and energy statistics, compute efficiency ratios, thermal and cooling means, and metrics on GPU nodes. These workload profiles describe the kind of workload behavior the segment represents. After these profiles are available, we partition them into workload archetypes using either K-Means or Hierarchical Density-Based Spatial Clustering of Applications with Noise (HDBSCAN), with the final choice between these two algorithms guided by the silhouette score~\cite{mcqueen1967some, campello2013density, rousseeuw1987silhouettes}. Clusters of fewer than 10 members are merged into the nearest centroid to avoid degenerate archetypes. This clustering enables anomaly scoring relative to similar workloads. We then fit an Isolation Forest separately within each cluster, which flags a segment as suspicious if it looks abnormal relative to other segments of the same workload archetype.

\subsubsection{Peer Analysis and Job Attribution} 
Cluster-conditioned scoring identifies segments that are unusual for a specific workload type. However, it does not directly exploit consistency among nodes executing the same job. When the same job runs across multiple nodes, those nodes should usually show comparable behavior. A node that drifts away from its peers is often more suspicious than the one whose absolute telemetry is simply high. To address this, we perform peer-group analysis for jobs spanning multiple nodes. For each multi-node segment, we build a running profile using telemetry accumulated up to the current minute. A node is flagged if its distance from the peer median is a robust outlier relative to the overall peer distribution, as measured by the Hampel Z-score:
\begin{equation}
z^{\mathrm{coh}}_i(T) = \frac{d_i(T) - \mathrm{median}_j\, d_j(T)}{k \cdot \mathrm{MAD}_j\, d_j(T)}.
\end{equation}
where $d_i(T)$ is the distance of node $i$ at time $T$, $\mathrm{median}_j d_j(T)$ and $\mathrm{MAD}_j d_j(T)$ are the median and median absolute deviation of these distances taken across all peer nodes $j$, and $k$ is the scaling constant that is used to make the MAD equivalent to a standard deviation for normally distributed data. A 30-minute warm-up precedes scoring, so that the running profile vector is stable enough for the peer median to be meaningful. We also apply a direction-aware per-minute check on system input power, because single-metric faults can be diluted in the full profile. This preserves cases in which a node draws substantially less or more power than its peers, which can indicate different failure modes.

Once the peer analysis is complete, we perform job attribution. When a single job occupies a segment, attribution is trivial because the running job is the only candidate. However, when two or more jobs share a node simultaneously, it is a challenging operational case because any of the concurrent jobs could cause the same anomalous power draw. We resolve this by scoring each candidate job across four signals: power residual, onset alignment, cluster deviation, and peer coherence. We combine these signals into a weighted attribution confidence score, with the weights redistributed proportionally when a signal is unavailable. The job with the highest score becomes the primary attributed job, and the full ranked candidate list is stored and surfaced later in the operator alert as a human-readable explanation that does not depend on the learned fusion stack.

Phase II, therefore, indicates whether the abnormal behavior is workload-related. The output of this stage is a set of job-aware signals, coherence scores, and the attribution evidence. These are then passed to Phase III, where the pipeline determines whether the same behavior is physically plausible.

\subsection{Phase III: Physics Reasoning}
Phase III tests whether anomalous observations are physically plausible by comparing telemetry with expected physical behavior. Physics-grounded constraints and a deterministic decision tree assign each detected anomaly a semantic label, producing interpretable per-minute evidence complementary to the statistical phases.

\subsubsection{Power and Thermal Model}
The power model uses a separate Ridge regression for each node type, trained across nodes on the training split. It predicts system input power from CPU load, memory usage, active job count, CPU package power, memory power, mean GPU utilization and the operating regime identifier. The expected power is modeled as:
\begin{equation}
    \hat{P}_h(t) = \beta_0+\beta^Tx_h(t)
\end{equation}
where $\hat{P}_h(t)$ is the predicted power for host $h$ at time $t$, $x_h(t)$ is the feature vector describing the node's workload and operating state, $\beta_0$ is the Ridge intercept, and $\beta$ is the Ridge coefficient vector. The model is trained across nodes to learn a general relationship between workload and expected power. Then it is adjusted with a host-specific bias correction to prevent false alarms due to idle behavior or hardware revisions. The final power residual is the difference between the observed power and the predicted power, plus the learned host-specific bias. Large positive or negative residuals indicate that a node is consuming more or less power than expected given its current workload.

The thermal model follows the same general idea as the power model. It estimates the expected inlet temperature using a similar Ridge regression technique. The model predicts the expected inlet temperature from system input power and the operating regime. This matters because thermal faults are often not obvious from absolute temperature values alone. The model captures the expected relationship between the power draw and the temperature response. A per-node calibration step is again applied because cooling efficiency varies across rack locations and hardware placement. The thermal residual is the difference between the observed and predicted inlet temperatures. In many cases, the important signal is that temperature changes are inconsistent with the amount of power being drawn, or that temperature rises without a corresponding workload or power driver.

\subsubsection{Adaptive Residual Scoring}
The residuals produced by the power and thermal models are usually not interpretable without a scale. We therefore use an adaptive residual scale rather than a fixed training-period standard deviation. This is important because HPC telemetry is non-stationary, and ambient conditions, workload mix, cooling behavior, and hardware state can shift over time without indicating a fault. For each residual channel, we estimate a robust rolling scale as:
\begin{equation}
    \sigma_{robust}(t) = \frac{IQR_{[t-14d, t]}(r)}{k}
\end{equation}
where $IQR_{[t-14d, t]}(r)$ is the interquartile range (IQR) of the residual $r$ calculated over a trailing 14-day window, and $k = 1.349$ converts the IQR to a standard-deviation scale. This rolling scale prevents future leakage while allowing the detector to adapt to slow operating-regime shifts. The residual scale is bounded between 0.5 and 5 times the training residual standard deviation. Residual flags require $|z| > 5$ for power or $|z| > 4$ for temperature. Around job starts and ends, we reduce the residuals by multiplying them by $1 - \exp(-\Delta/\tau)$, where $\Delta$ is the time to the nearest job start or end and $\tau = 5\,\mathrm{min}$. This adjustment applies within 20 minutes of a transition to avoid treating normal startup and shutdown behavior as faults.

\subsubsection{Constraint Rules}
Residuals alone, however, are not sufficient to explain what kind of inconsistency is present. We therefore apply a small library of physics-grounded constraint rules. These rules are meant to check structural relationships that should hold even when no single sensor value is extreme. We apply \textit{four} primary constraint rules to audit the physical consistency of the data. The \textbf{dynamics constraint} ensures that temperature changes are reflected in power consumption. It helps identify anomalies such as increased heat on idle nodes. A \textbf{cross-plane constraint} compares rack-level PDU measurements with the aggregate of individual BMC data in that rack. The discrepancies typically indicate sensor drift or reporting failures. The \textbf{allocation-versus-utilization constraint} identifies nodes officially assigned to a job but with no measurable activity. The \textbf{cooling-response constraint} identifies nodes whose inlet temperature climbs while fan speed fails to respond. To ensure result integrity, each constraint requires a sustained run of at least three to five consecutive minutes before firing, with the allocation-versus-utilization rule requiring 30 minutes to reflect the natural idle gaps within legitimate jobs.

Therefore, Phase III combines residual magnitude, constraint evidence, and context weights. This stage helps suppress deviations explained by normal workload behavior, while also highlighting deviations without a meaningful explanation. The full semantic label is passed to Phase IV to generate operator-facing explanations.

\subsection{Phase IV: Fusion}
Phase IV receives node-local statistical signals from Phase I, workload- and peer-aware signals from Phase II, and physics residuals, constraint violations, and semantic labels from Phase III. It fuses this heterogeneous evidence into a single episode-level decision. Phases I-III are fitted without anomaly-classification targets, but known labeled fault intervals are excluded from training alongside maintenance periods and audit-flagged readings. Their thresholds are not tuned against event labels. Phase IV uses labeled episodes for \textit{supervised} fusion, with a predefined feature set and validation-selected tier thresholds. Test labels do not enter fitting, feature selection or threshold tuning.

\subsubsection{Episode Formation}
\label{episode}
Although the upstream phases produce per-minute evidence, reporting decisions at minute granularity are not well aligned with operational practice. HPC operators typically investigate incidents as sustained episodes involving nodes, jobs, and subsystems rather than as isolated timestamps. Moreover, excessive undifferentiated alerts can reduce trust in the detector and cause important events to be overlooked~\cite{voutsas2024mitigating}. Therefore, Phase IV first groups per-minute evidence into candidate anomaly episodes before performing the final classification.

A candidate episode is formed when a phase headline score is sustained above a small threshold for several consecutive minutes. Adjacent candidates separated by short gaps are merged so that intermittent or flickering faults are treated as a single event. This episode-level formulation also matches the available ground truth, since operator tickets and node-failure records usually describe time intervals rather than individual timestamps. For each candidate episode, we condense the upstream evidence into a fixed-length feature vector. Instead of using only one summary scalar per phase, we retain the raw detector flags and supporting signals, allowing the fusion model to learn cross-phase interactions, such as the concurrence of statistical consensus in Phase I and a physics violation in Phase III. Each episode also captures context unavailable to per-minute detectors: how its evidence ranks against the host's own recent history, how persistently each phase fired rather than only how strongly, how many hosts in the same rack had an episode open concurrently, how often that host has recently produced weak candidates, and how its evidence compares with that of its job peers.

\subsubsection{Gradient Boosting Classifier}
After the episodes are formed, we train a histogram-based gradient-boosting classifier on episode-level features to produce a fused anomaly probability. The fusion model learns which combinations of statistical, workload-aware, and physics-guided evidence are most predictive of true anomalies. The classifier output is mapped into three operating tiers: \texttt{CRITICAL}, \texttt{CONFIRMED}, and \texttt{CANDIDATE}. The \texttt{CRITICAL} tier emphasizes high precision, the \texttt{CONFIRMED} tier serves as the primary reporting point, and the \texttt{CANDIDATE} tier prioritizes higher recall. This tiered design lets operators choose an operating point based on the cost of missed events versus the cost of unnecessary interruptions. All three thresholds are selected on the validation split alone and are never tuned on test, and we report the validation-to-test transfer of that policy rather than a test-optimal point.

To make the final output actionable, each alert is emitted as a compact explanation package rather than a numerical score alone. The package includes the fused probability, tier assignment, semantic context label, dominant contributing signals, and TreeSHAP attribution from the fusion model. When available, it also includes fired rules, violated constraints, and implicated job context. These explanations give operators a starting point for triage and make the decision auditable. Thus, Phase IV serves as the final decision layer that transforms per-minute detector evidence into scored, tiered, and interpretable anomaly episodes.

\section{Evaluation and Results}
In this section, we provide a detailed overview of the experimental setup, the controlled fault injections, the baseline methods, and the overall performance of \projectname.

\subsection{Experimental Setup}
We use an NSF-funded REPACSS cluster, comprising 110 CPU nodes, 8 GPU nodes, 3 login nodes, and 9 storage nodes, totaling 28,984 cores and 176 TB of system memory. Each CPU node is equipped with two AMD EPYC 9754 processors providing 256 cores and 1.5 TB of DDR5 memory per node. Each GPU node contains two Intel Xeon Gold 6448Y processors and four NVIDIA H100 NVL GPUs interconnected via NVLink, each with 94 GB of HBM memory. The cluster is monitored through a centralized monitoring service backed by TimescaleDB. It ingests metrics from multiple sources, including BMC and PDU metrics and Slurm workload data.

\begin{figure}[t]
\centering
\includegraphics[width=\columnwidth]{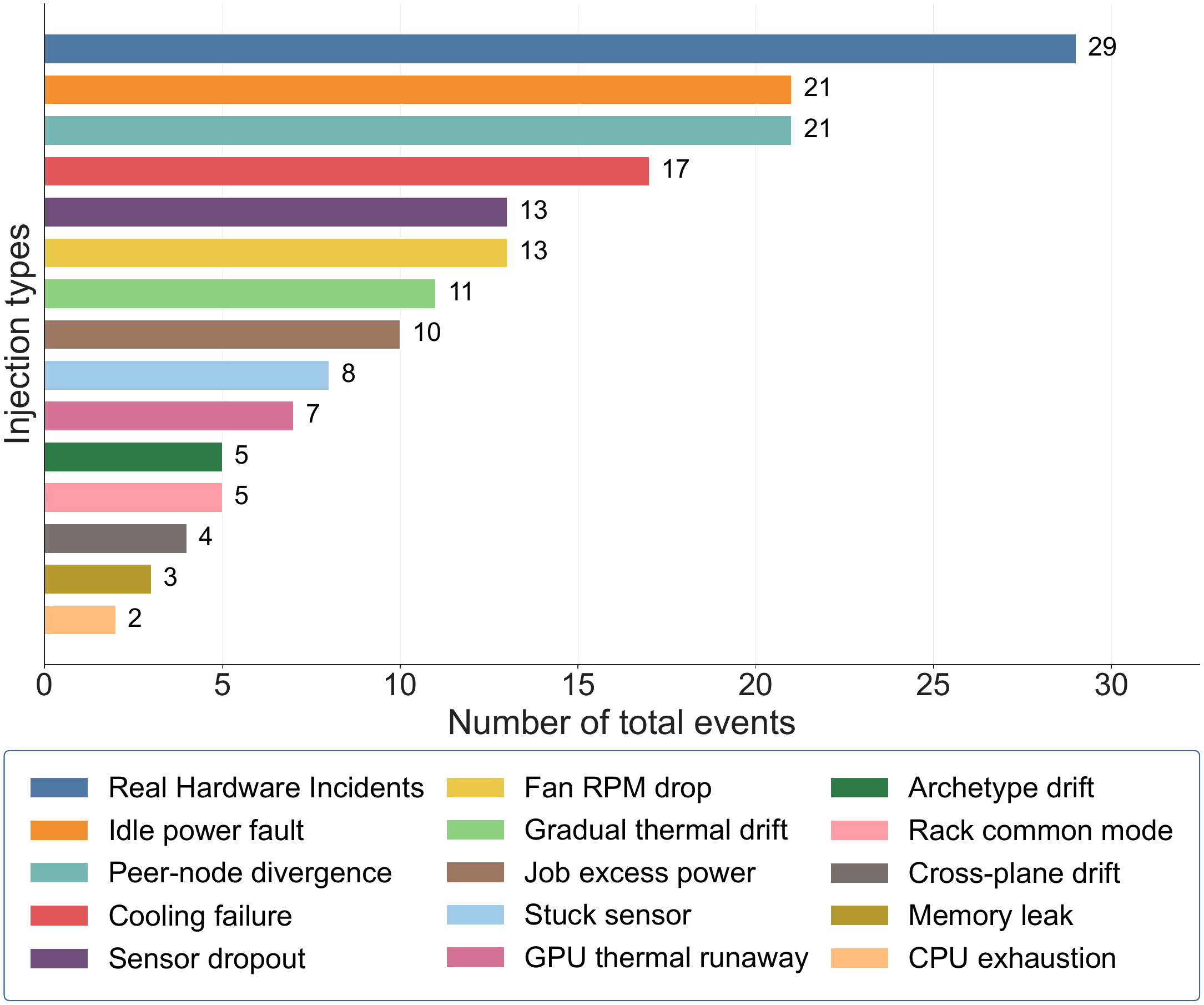}
\caption{Types of anomalies in the test split of the dataset.}
\label{fig:injected_anomalies}
\end{figure}

\subsection{Dataset and Controlled Fault Injections}
We evaluate \projectname~on 79 days of heterogeneous production telemetry from the 118-node compute fleet, spanning January 1 to March 20, 2026. The dataset covers approximately 220K node-hours of monitoring and 13,266 Slurm jobs, integrating node, workload, and infrastructure telemetry. The first 59 days form the training set, the next 8 days form the validation set, and the final 12 days form the test set. A maintenance shutdown between February 14 and February 16, 2026, is masked from the training window.

Ground-truth failures are sparse and often recorded only after operator investigation~\cite{das2023prolego}. Because naturally occurring labeled failures provide limited coverage of the diverse failure modes encountered in HPC systems, controlled injections allow us to evaluate detector behavior across a broader and systematically defined set of operational conditions. We therefore evaluate on 169 labeled test events across 85 hosts: 29 operator-recorded failures and 140 controlled fault injections spanning HPC-relevant thermal, power, cooling, resource, sensing, and cross-plane failure modes, as shown in Figure~\ref{fig:injected_anomalies}. Idle-power injections add 110–220 W for 2–6 h, cooling injections raise inlet temperature by 4–6 °C for 1–2 h, and sensor dropouts last 2–3 h. These test selected signatures rather than the full diversity of production failures. Training and validation use independently generated events, and no test-window label is used for training, threshold selection, or model selection. 


\subsection{Baseline Methods}
We compare \projectname~against four established unsupervised baselines: two published state-of-the-art anomaly detection systems, Prodigy~\cite{aksar2023prodigy} and NodeSentry~\cite{xia2025effective}, and two standard unsupervised detectors, Isolation Forest~\cite{liu2008isolation} and LSTM Autoencoder~\cite{malhotra2016lstm}. All baselines are trained using the same 59-day training window and evaluated on the same temporal validation and test splits as \projectname. Prodigy uses healthy training data and a 99th-percentile reconstruction-error threshold. NodeSentry is fitted without labels, but its threshold maximizes validation F1. Isolation Forest and LSTM-AE use Phase I's training data and their native anomaly flags.

\subsection{Evaluation Metrics}
We report precision, recall, F1, AUROC, and AUPRC, together with event recall. Phases I-III are fitted on the training data. Phase IV is fitted on labelled training and validation episodes, with validation labels also used to select its tier thresholds. Final evaluation uses the held-out test split. To absorb both imprecision in operator-recorded fault boundaries and the alignment granularity of windowed scoring, each ground-truth event is expanded into a tolerance window extending 60 minutes before its start and 30 minutes after its end. A candidate episode matches a ground-truth event if it overlaps the event's tolerance window on the same host, on any host in the same rack for rack- or PDU-level events, or on any host for cluster-wide events. AUROC/AUPRC rank method-specific pools. Each alert episode matching at least one ground-truth tolerance window counts as one true positive (TP). An unmatched alert counts as one false positive (FP). Each event with no matching alert counts as one false negative (FN). Reported recall is $\mathrm{TP}/(\mathrm{TP}+\mathrm{FN})$, combining matched alert episodes with missed events. We separately report event recall, the fraction of events matched by at least one alert.
\begin{table}[t]
\centering
\caption{Performance comparison of \textsc{\projectname} and baselines.}
\label{tab:performance_metrics}
\footnotesize

\resizebox{\columnwidth}{!}{%
\begin{tabular}{lccccc}
\toprule
\textbf{Method} & \textbf{Prec.} & \textbf{Rec.} & \textbf{F1} & \textbf{AUROC} & \textbf{AUPRC} \\
\midrule
\textbf{\textsc{\projectname}} & \textbf{0.913} & \textbf{0.817} & \textbf{0.862} & \textbf{0.896} & \textbf{0.878} \\
NodeSentry        & 0.710 & 0.452 & 0.552 & 0.379 & 0.489 \\
Prodigy           & 0.599 & 0.735 & 0.660 & 0.666 & 0.450 \\
Isolation Forest  & 0.436 & 0.560 & 0.490 & 0.811 & 0.760 \\
LSTM-AE           & 0.491 & 0.470 & 0.480 & 0.728 & 0.758 \\
\bottomrule
\end{tabular}%
}
\end{table}


\subsection{Overall Performance}
Table~\ref{tab:performance_metrics} compares \projectname~with the baseline methods. At the validation-selected \texttt{CRITICAL}, \texttt{CONFIRMED}, and \texttt{CANDIDATE} thresholds, precision/recall are 0.979/0.525, 0.913/0.817, and 0.813/0.899, respectively. At the primary \texttt{CONFIRMED} operating point, \projectname~achieves the best overall performance, reaching an F1-score of 0.862 with a precision of 0.913 and a recall of 0.817, with an event recall of 0.805. This represents a 31\% improvement in F1 over the next strongest baseline, showing that workload-aware and physics-guided evidence adds value beyond statistical anomaly scores alone. As a sensitivity check, re-evaluating the frozen alerts with a stricter 0/15-minute tolerance instead of 60/30 minutes yields nearly unchanged precision/recall/F1 of 0.907/0.811/0.856. NodeSentry’s AUROC is low because many episodes during the cooling incident receive lower scores than GPU episodes with no matching recorded fault.

\subsection{Computational Cost and Scalability}
All training and inference run on a single 20-core allocation with one NVIDIA H100 GPU. The end-to-end replay pipeline, from preprocessing through training and evaluation of all four phases, over 79 days of telemetry from 118 nodes, completes in approximately 43 minutes of wall-clock time. Measured on node-limited replays of the raw telemetry at 38, 68, and 98 nodes, runtime scales nearly linearly with fleet size at 14.6 s per node plus an 850 s fixed cost ($R^2=0.997$). A linear extrapolation of this fit predicts approximately 4.3 hours for 1,000 nodes on the same configuration. Frozen-model inference also scales linearly at 15.1 s per node ($R^2=0.999$) for the same 79-day dataset, which is about 0.13 ms of compute per node-minute of telemetry. These timings measure throughput rather than alert delay. Classification uses features summarized over complete episodes.

\begin{table}[t]
\centering
\caption{Performance comparison of \textsc{\projectname} and different components.}
\label{tab:ablation_metrics}
\footnotesize

\resizebox{\columnwidth}{!}{%
\begin{tabular}{lccccc}
\toprule
\textbf{Method} & \textbf{Prec.} & \textbf{Rec.} & \textbf{F1} & \textbf{AUROC} & \textbf{AUPRC} \\
\midrule
\textbf{\textsc{\projectname}} & \textbf{0.913} & 0.817 & \textbf{0.862} & \textbf{0.896} & 0.878 \\
P-I                                                   & 0.393 & 0.955          & 0.557 & 0.803 & 0.785 \\
P-II                                                  & 0.692 & 0.178          & 0.283 & 0.849 & \textbf{0.940} \\
P-III                                                 & 0.361 & 0.954         & 0.524 & 0.729 & 0.705 \\
$\text{P-I} \wedge \text{P-II}$                       & 0.550 & 0.203         & 0.296 & 0.855 & 0.893 \\
$\text{P-I} \wedge \text{P-II} \wedge \text{P-III}$   & 0.571 & 0.171         & 0.263 & 0.758 & 0.842 \\
$\text{P-I} \lor \text{P-II} \lor \text{P-III}$       & 0.318 & \textbf{0.988} & 0.481 & 0.734 & 0.689 \\
Majority vote ($\geq 2$)                              & 0.537 & 0.911          & 0.675 & 0.811 & 0.862 \\
\bottomrule
\end{tabular}%
}
\end{table}

\begin{table}[t]
\centering
\caption{Eclipse cross-dataset evaluation.}
\label{tab:eclipse_eval}
\footnotesize

\resizebox{\columnwidth}{!}{%
\begin{tabular}{c|l|cccc}
\toprule
\textbf{Task} & \textbf{Method} & \textbf{Prec.} & \textbf{Rec.} & \textbf{F1} & \textbf{AUROC} \\
\midrule
\multirow{3}{*}{\textbf{memleak}}
  & \textbf{\textsc{\projectname}}     & 0.859          & 0.994 & 0.921 & 0.735 \\
  & $\text{Prodigy}$                  & \textbf{0.940} & \textbf{1.000} & \textbf{0.969} & \textbf{1.000} \\
  & $\text{NodeSentry}$               & 0.845          & 0.982          & 0.908          & 0.683 \\
\midrule
\multirow{3}{*}{\textbf{cpuoccupy}}
  & \textbf{\textsc{\projectname}}     & 0.852          & 0.971          & 0.907          & 0.692 \\
  & $\text{Prodigy}$                  & \textbf{0.864} & \textbf{1.000} & \textbf{0.927} & \textbf{0.695} \\
  & $\text{NodeSentry}$               & 0.852          & 0.965          & 0.904          & 0.689 \\
\bottomrule
\end{tabular}%
}
\end{table}

\subsection{Ablation Study}
The contribution of each design choice in \projectname~is shown in Table~\ref{tab:ablation_metrics}. All configurations use the same upstream features and event-matching protocol. The ablation compares \projectname~against several non-learned fusion baselines that combine phase outputs using fixed Boolean logic. The OR operator ($\lor$) reports an episode if any phase flags it as anomalous. The AND operator ($\wedge$) reports an episode only when all required phases agree. Majority voting requires agreement from two or more phase-level detectors. \projectname~ranks the shared Phase IV candidate pool, including candidates below alert thresholds. Isolation Forest, LSTM-AE and Boolean configurations rank flagged episodes. Prodigy ranks scored windows. NodeSentry forms episodes from at least two consecutive minutes above each host’s 75th-percentile score, merging gaps shorter than 10 minutes. It ranks them by maximum reconstruction error. AUROC/AUPRC are therefore not directly comparable across rows. The results show that neither individual phases nor fixed Boolean fusion is sufficient. P-II's high AUPRC reflects a small, high-base-rate pool rather than stronger detection, as its recall of 0.178 shows. \projectname~achieves the strongest overall precision-recall balance, while maintaining an F1-score of 0.862. Learned fusion outperforms the tested Boolean rules. These comparisons do not isolate each phase’s contribution under a retrained classifier.

\subsection{Portability Beyond BMC Telemetry}
To test portability beyond BMC telemetry, we evaluate \projectname~on the Eclipse dataset used by Prodigy~\cite{aksar2023prodigy}. This dataset provides 240 labeled job-node pairs across four HPC applications and two anomaly types (memleak and cpuoccupy), which are evaluated as two binary tasks of 160 samples each following Prodigy's protocol. Its telemetry differs substantially from ours: 156 Linux $/proc$ counters at 1 Hz, without temperature, fan, power, or useful peer-divergence signals. The traces are also pre-sliced into 22–56-minute pairs rather than collected as continuous multi-week streams. This setup lets us compare against Prodigy on its native benchmark and against NodeSentry~\cite{xia2025effective} under the same protocol. Under this protocol, only Phase I of our model is exercised. Phase II is disabled under this protocol because its peer-divergence signal is uninformative when the anomaly is applied to all four nodes of a job, and Phase III is not applicable because the dataset lacks the physical signals required by this phase. Our model therefore reduces to Phase I's Isolation Forest and LSTM-Autoencoder stack, ported without code modification. For a fair comparison, we followed Prodigy's evaluation protocol, which includes three random seeds and an 80/20 split for the healthy training pool. We tuned our thresholds on a disjoint 30\% validation set and reserved the remaining 70\% for the final test phase. NodeSentry uses four clusters, matching Eclipse's four applications, with the same splits. 

With only Phase I available, \projectname~performs similarly to Prodigy, as shown in Table~\ref{tab:eclipse_eval}. It trails by only a small margin on memleak (0.921 vs. 0.969 F1) and cpuoccupy (0.907 vs. 0.927), while still outperforming NodeSentry.

\subsection{Case Studies}

\begin{figure}[t]
\centering
\includegraphics[width=\columnwidth]{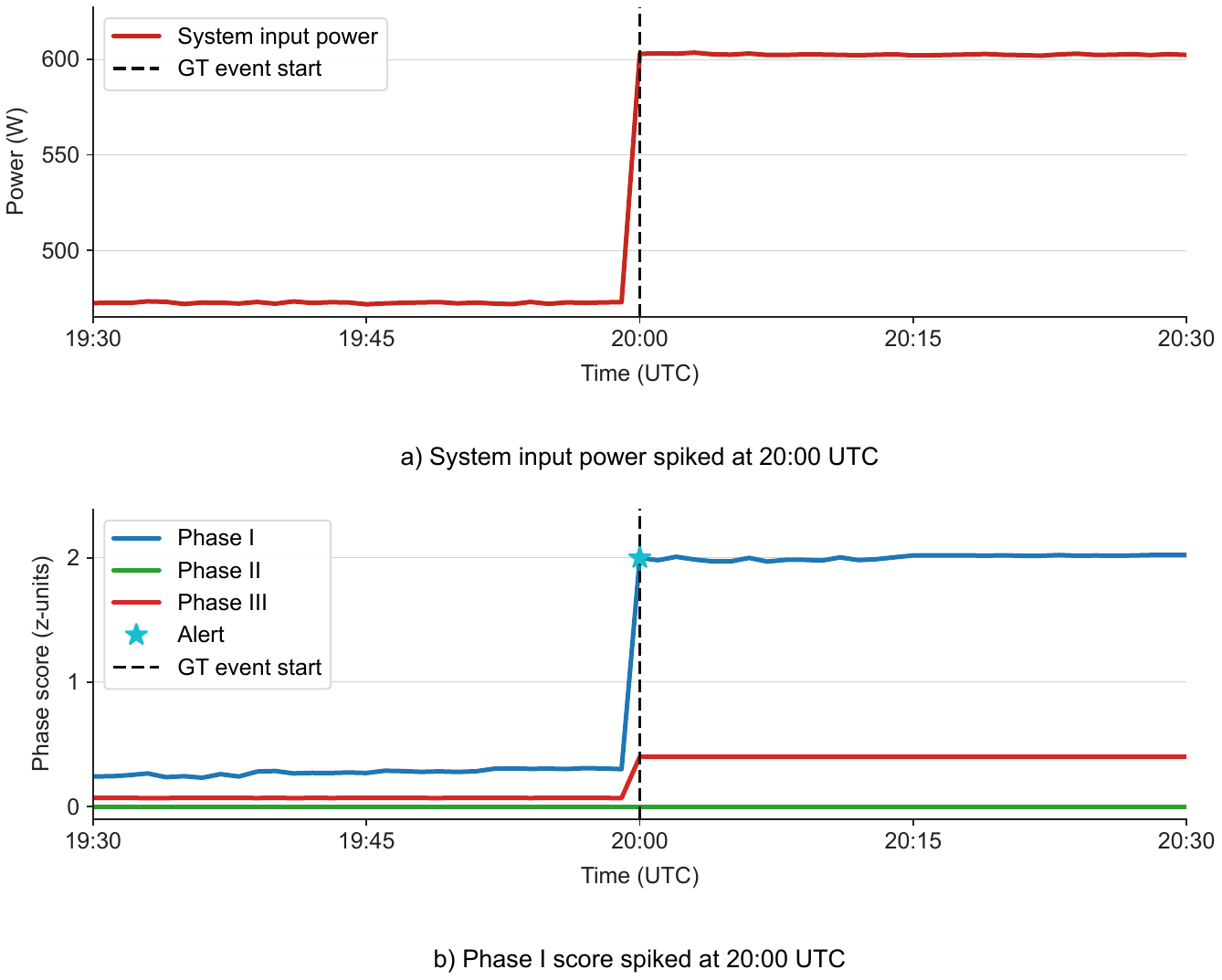}
\caption{An input-power spike triggered Phase I; Phase III validated it.}
\label{fig:caseA}
\end{figure}













\definecolor{alertred}{RGB}{165,42,42}
\definecolor{alertgray}{RGB}{245,245,245}
\definecolor{alertorange}{RGB}{217,130,43}
\definecolor{darkgray}{RGB}{70,70,70}

\begin{figure}[t]
\centering
\scriptsize


\begin{tcolorbox}[
    width=0.96\columnwidth,
    colback=white,
    colframe=black!60,
    boxrule=0.45pt,
    arc=1pt,
    left=6pt,
    right=6pt,
    top=5pt,
    bottom=5pt,
    title=\textbf{Explainable Alert: Idle Node Power Excess},
    halign title=center,
    coltitle=black,
    colbacktitle=black!8,
    fonttitle=\scriptsize\bfseries
]

\textbf{Target:} CPU node rpc-97-1 \hfill
\textbf{Severity:} \textcolor{alertred!85!black}{\textbf{CRITICAL}}

\vspace{3pt}

\begin{tcolorbox}[
    colback=alertgray,
    colframe=black!20,
    boxrule=0.25pt,
    arc=1pt,
    left=4pt,
    right=4pt,
    top=3pt,
    bottom=3pt
]
\begin{tabular}{@{}ll@{}}
\textbf{Detection phase:} & Phase I, supported by Phase III \\
\textbf{Episode onset:}    & 2026-03-10 20:00 UTC \\
\textbf{Associated job:} & None \\
\textbf{Likely cause:}   & Parasitic process or firmware issue \\
\end{tabular}
\end{tcolorbox}

\vspace{3pt}

\textbf{Operator action.}
Investigate within 15 min. CPU node rpc-97-1 drew sustained power above its 7-day idle baseline despite having no active job. Schedule the node for reboot if no legitimate background activity is identified.

\vspace{4pt}

\textbf{Supporting evidence.}
\begin{itemize}[leftmargin=1.2em,itemsep=1pt,topsep=1pt]
    \item System input power moved from a baseline of 472.3 W to 602.7 W (+130.4 W, +28\% of baseline).
    \item Phase III physics residual reached 9.1$\sigma$, outside the model-expected envelope for this node.
    \item Phase I rule that fired: Idle Power Surge
\end{itemize}

\end{tcolorbox}

\caption{Explainable alert generated by \textsc{\projectname}.}
\label{fig:explainable_alert}
\end{figure}

\subsubsection{Idle Power Fault} \projectname~raised a \texttt{CRITICAL}-tier alert on a CPU node rpc-97-1. Figure~\ref{fig:caseA} shows the anomaly onset. The 240-minute injection (a sustained $+130 \mathrm{W}$ draw on an idle node with no Slurm allocation, 2026-03-10 20:00 UTC) was bracketed minute-for-minute by a 240-minute alert. The node remained idle for the preceding 24 hours. Phase I's IDLE\_POWER\_SURGE rule and Phase III’s physics evidence both begin at the fault onset in replay. Note that Figure~\ref{fig:caseA}b plots each phase's bounded episode-formation score in z-units, which compresses large residuals. Figure~\ref{fig:explainable_alert} shows the operator-facing alert card.

\begin{figure}[t]
\centering
\includegraphics[width=\columnwidth]{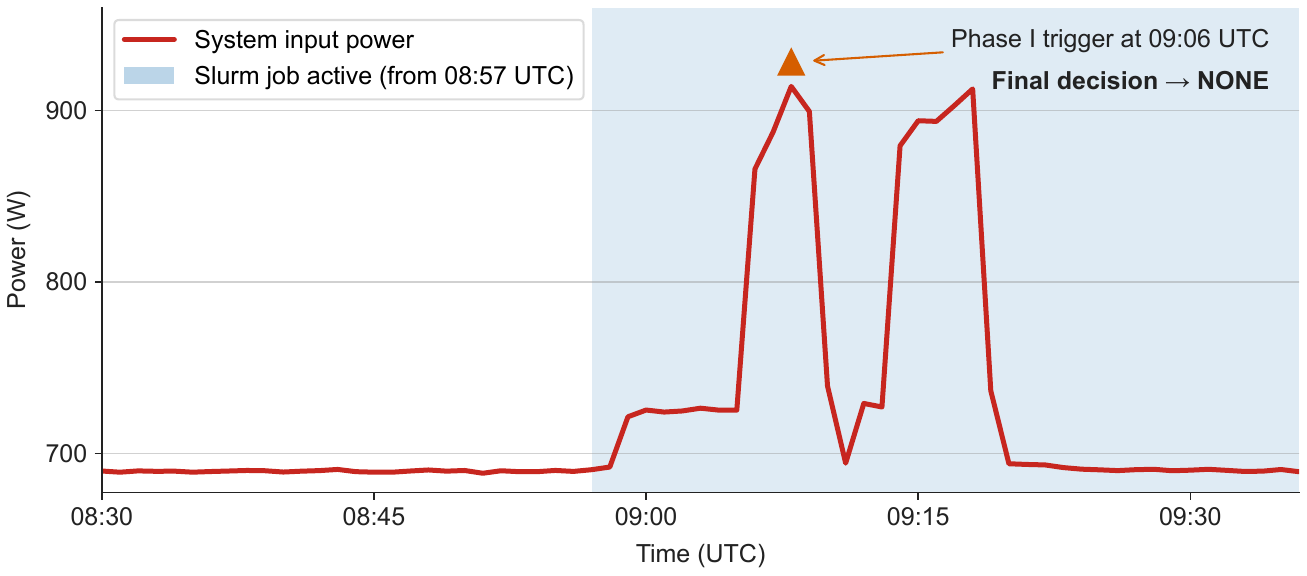}
\caption{A workload-explained power burst on GPU node rpg-93-7.}
\label{fig:caseB}
\end{figure}

\subsubsection{False Positive Rescue} During training-period replay, a Slurm job started at 08:57 UTC on 2026-01-04 on GPU node rpg-93-7 and drove system input power from a 690 W idle baseline to bursts near 915 W, as shown in Figure~\ref{fig:caseB}. It caused Phase I to spike to 2.6$\sigma$. However, Phase II found the workload behavior routine and Phase III found no physics violation ($|z|\leq1.3$), so the fusion stage suppressed the episode to \texttt{NONE} and no operator alert was issued.

\section{Conclusion and Future Work}
In this paper, we presented \projectname, a multi-plane anomaly-detection architecture for HPC telemetry. \projectname~progressively refines statistical anomaly evidence using workload context, peer behavior, and physical consistency before episode-level fusion. Across 79 days of production telemetry with 29 operator-recorded failures and 140 controlled fault injections, it achieved 0.913 precision, 0.817 recall, and 0.862 F1 while producing auditable operator-facing alerts. The present evaluation is retrospective. Future work will study long-term online deployment and adaptation under evolving workloads.

\section{Acknowledgment}
We are thankful to the anonymous reviewers for their valuable feedback. This research is supported in part by the National Science Foundation under grants OAC-2404438 and CNS-1939140 (A U.S. National Science Foundation Industry-University Cooperative Research Center on Cloud and Autonomic Computing). We used ChatGPT-5.5 to assist with improving grammar and sentence clarity. All ideas and content remain original.

\bibliographystyle{IEEEtran}
\bibliography{references}

\end{document}